\documentclass[a4paper,numcites]{aipproc}
\layoutstyle{6s}

\usepackage{latexsym}
\usepackage{revsymb}
\usepackage{amsfonts}
\usepackage{amsmath}
\usepackage{amssymb}
\usepackage{bm}

\newcommand{\vev}[1]{\langle #1\rangle}
\newcommand{\bra}[1]{\langle #1 |}
\newcommand{\ket}[1]{| #1 \rangle}
\newcommand{\braket}[2]{\langle #1 | #2\rangle}

\newcommand{\0}{\underline{0}}
\newcommand{\1}{\underline{1}}

\renewcommand{\k}{\underline{k}}
\renewcommand{\l}{\underline{\ell}}

\newcommand{\K}{\mathbb{K}}

\newcommand{\C}{\mathbb{C}}
\newcommand{\R}{\mathbb{R}}
\newcommand{\N}{\mathbb{N}}
\newcommand{\Z}{\mathbb{Z}}

\begin{document}

\title{Some Mutant Forms of Quantum Mechanics}



\author{Tatsu Takeuchi\footnote{Presenting author}}{address={Department of Physics, Virginia Tech, Blacksburg, VA 24061, USA},email={takeuchi@vt.edu}}
\author{Lay Nam Chang}{address={Department of Physics, Virginia Tech, Blacksburg, VA 24061, USA},email={laynam@vt.edu}}
\author{Zachary Lewis}{address={Department of Physics, Virginia Tech, Blacksburg, VA 24061, USA},email={zlewis@vt.edu}}
\author{Djordje Minic}{address={Department of Physics, Virginia Tech, Blacksburg, VA 24061, USA},email={dminic@vt.edu}}

\begin{abstract}
We construct a `mutant' form of quantum mechanics on a vector space over 
the finite Galois field $GF(q)$.
We find that the correlations in our model do not violate the
Clauser-Horne-Shimony-Holt (CHSH) version of Bell's inequality,
despite the fact that the predictions of this discretized quantum mechanics 
cannot be reproduced with any hidden variable theory.
An alternative `mutation' is also suggested.
\end{abstract}

\maketitle

Among many features that distinguish canonical quantum mechanics (QM) 
from its classical counterpart are the super-classical correlations that violate
Bell's inequality \cite{Bell:1964kc,BellBook}.
In the formulation of Clauser, Horne, Shimony, and Holt (CHSH) \cite{Clauser:1969ny},
the inequality is expressed as 
a bound on the absolute value of the following combination of correlators:
\begin{equation}
\vev{A,a;B,b}\;\equiv\;\vev{AB}+\vev{Ab}+\vev{aB}-\vev{ab}\;.
\end{equation}
Here, $A$ and $a$ are observables of particle 1, while $B$ and $b$ are observables of 
particle 2.  All four observables are assumed to yield either $+1$ or $-1$ upon measurement.
Since each of the four correlators that contribute to this combination must have
a value in the range $[-1,1]$, the absolute maximum value that $|\vev{A,a;B,b}|$ can have is $4$.
However, neither classical hidden variable theories nor canonical QM can saturate this bound;
for classical hidden variable theories, the upper bound on $|\vev{A,a;B,b}|$ is $2$,
while that for canonical QM is the Cirel'son value of $2\sqrt{2}$ \cite{cirelson,landau}.
Thus, the CHSH bound not only distinguishes between classical theories and canonical QM,
but it also distinguishes between canonical QM and possible super-quantum theories \cite{super,Chang:2011yt}.

In order to illuminate which mathematical aspect of canonical QM 
is responsible for its specific physical features,
we adopt what we call the `geneticist's approach.'  
We introduce `mutations' into QM by knocking out some of its mathematical `genes'
and investigate which of the physical characteristics of canonical QM survives the `mutation'
and which ones do not.
In particular, we look at how the value of the CHSH bound changes with each mutation.
Through such studies, we hope to clarify the connections between
the mathematical genotype and the physical phenotype.

The `mutation' we introduce in this talk is the replacement of the
number field over which the vector space of states is defined.  
In canonical QM, this is of course the field of complex numbers $\C$.
We replace this with the finite Galois field $GF(q)$, where $q=p^n$, $n\in\N$, and
$p$ is a prime number.  For the $n=1$ case, they are $GF(p)=\Z_p=\Z/p\Z$.
A similar proposal to replace $\C$ with $GF(q)$ has been made previously 
by Schumacher and Westmoreland in Ref.~\cite{MQT},
in which probabilities were not defined.

Using the shorthand $\K=GF(q)$, the state space for an $N$-level system 
in our approach is $V=\K^N$.
This space lacks an inner product, and consequently, also lacks normalizable states, 
symmetric/hermitian operators, and the usual pairing of vectors in $V$ with dual vectors
in $V^*$ via the inner product.
Nevertheless, we can construct a quantum-like theory on it as follows.

First, the states of the system are represented by vectors $\ket{\psi}\in V$,
while the outcomes of measurements are represented by dual-vectors $\bra{x}\in V^*$.
An observable is associated with a choice of basis in $V^*$, with each dual-vector
in the basis representing a different outcome.
The probability of obtaining the outcome represented by $\bra{x}$ when the
observable is measured on the state represented by $\ket{\psi}$ is assumed to be given by
\begin{equation}
P(x|\psi) \;=\; \dfrac{\bigl|\braket{x}{\psi}\bigr|^2}{\sum_y \bigl|\braket{y}{\psi}\bigr|^2}\;,
\label{Pdef}
\end{equation}
where the sum in the denominator runs over all the dual-vectors in the basis, and
the absolute value function is defined as
\begin{equation}
|\,\underline{k}\,|\;=\;
\left\{\begin{array}{ll}
0\quad &\mbox{if $\underline{k}=\0$}\;,\\
1\quad &\mbox{if $\underline{k}\neq\0$}\;.
\end{array}
\right.
\label{abs}
\end{equation}
Note that this function is product preserving, that is, $|\k\l|=|\k||\l|$ for all $\k,\l\in\K$.
This assignment gives equal probabilistic weight to all outcomes $\bra{x}$ for which $\braket{x}{\psi}\neq\0$.
It also leads to 
the vectors $\ket{\psi}$ and $\k\ket{\psi}$, where $\k\in\K\backslash\{\0\}$,
having the exact same probabilities for all outcomes $\bra{x}$.
Thus, vectors that differ by non-zero multiplicative constants can be identified
as representing the same physical state.
This endows the state space of the model with the
finite projective geometry \cite{Hirschfeld}
\begin{eqnarray}
PG(N-1,q) 
\;=\; (\,\K^N\backslash\{\mathbf{\0}\}\,)\,\big/\,(\,\K\backslash\{\0\}\,) 
\;,
\end{eqnarray}
where each `line' going through the origin of $V=\K^N$ is identified as a `point,'
in close analogy to the complex projective geometry of the state space of canonical QM:
\begin{equation}
\mathbb{C}P^{N-1}
\;=\; (\,\mathbb{C}^{N}\backslash\{\mathbf{0}\}\,)\,\big/\,(\,\mathbb{C}\backslash\{0\}\,)
\;.
\end{equation}

To give a concrete example of our proposal, let us 
set $N=2$ and $\K=GF(2)=\Z_2=\{\0,\1\}$.
In this case, there are only three states in $V=\Z_2^2$ :
\begin{equation}
\ket{\,a\,} = \left[\begin{array}{c} \1 \\ \0 \end{array}\right],\qquad
\ket{\,b\,} = \left[\begin{array}{c} \0 \\ \1 \end{array}\right],\qquad
\ket{\,c\,} = \left[\begin{array}{c} \1 \\ \1 \end{array}\right],
\end{equation}
and three outcomes in $V^*$ :
\begin{equation}
\bra{\,\overline{a}\,} = \bigl[\;\0\;\;\1\;\bigr]\;,\qquad
\bra{\,\overline{b}\,} = \bigl[\;\1\;\;\0\;\bigr]\;,\qquad
\bra{\,\overline{c}\,} = \bigl[\;\1\;\;\1\;\bigr]\;.
\end{equation}
We have labeled the vectors and dual-vectors so that the brackets are given by
\begin{equation}
\braket{\bar{r}}{s} 
\;=\;
\begin{cases}
\0 & \mbox{if $r=s$}\;, \\
\1 & \mbox{if $r\neq s$}\;,
\end{cases}
\end{equation}
and their absolute values are
\begin{equation}
\bigl|\braket{\bar{r}}{s}\bigr|\;=\; 1-\delta_{rs}\;.
\label{braketpq}
\end{equation}
Observables are associated with a choice of basis of $V^*$:
\begin{equation}
A_{rs}\;\equiv\;\{\;\bra{\bar{r}},\;\bra{\bar{s}}\;\}\;,\qquad
r\neq s\;.
\end{equation}
We assign the outcome $+1$ to the first dual-vector of the pair, 
and the outcome $-1$ to the second to make these observables
spin-like. This assignment implies $A_{sr}=-A_{rs}$.
The indices $rs$ can be considered as indicating the direction of the `spin,'
and the interchange of the indices as indicating a reversal of this direction.
In the current case, there are three independent `spins,' which we denote
\begin{equation}
Z\;=\;A_{ab}\;=\;\{\bra{\bar{a}},\bra{\bar{b}}\}\;,\quad
X\;=\;A_{bc}\;=\;\{\bra{\bar{b}},\bra{\bar{c}}\}\;,\quad
Y\;=\;A_{ca}\;=\;\{\bra{\bar{c}},\bra{\bar{a}}\}\;.
\end{equation}
Applying Eq.~\eqref{Pdef} to this system, it is straightforward to show that
\begin{eqnarray}
& &
P(Z=+1\,|\,a) \;=\; 0\;,\quad
P(Z=+1\,|\,b) \;=\; 1\;,\quad
P(Z=\pm 1\,|\,c) \;=\; \frac{1}{2}\;,
\cr
& &
P(Z=-1\,|\,a) \;=\; 1\;,\quad
P(Z=-1\,|\,b) \;=\; 0\;,
\end{eqnarray}
and thus,
\begin{equation}
\vev{Z}_a \,=\, -1\;,\qquad
\vev{Z}_b \,=\, +1\;,\qquad
\vev{Z}_c \,=\, \phantom{-}0\;.
\end{equation}
The probabilities and expectation values for $X$ and $Y$ can be obtained by 
simultaneous cyclic permutations of the labels $(XYZ)$ and $(abc)$.
Thus for each `spin,' there exist two `eigenstates,'
one for $+1$ (`spin' up) and another for $-1$ (`spin' down).
For the third state the two outcomes $\pm 1$ are equally probable.

Two particle `spin' states are constructed on the tensored space 
$V\otimes V = \Z_2^2\otimes\Z_2^2 = \Z_2^4$.
There are $2^4-1=15$ non-zero vectors in this space, 
of which $3^2=9$ are product states.  The remaining
$15-9=6$ are entangled and are given by:
\begin{eqnarray}
\ket{S}
& = & \ket{a}\otimes\ket{a}+\ket{b}\otimes\ket{b}+\ket{c}\otimes\ket{c}
\;=\; \bigl[\;\0\;\;\1\;\;\1\;\;\0\;\bigr]^\mathrm{T} \;,\cr
\ket{(ab)} 
& = & \ket{a}\otimes\ket{b}+\ket{b}\otimes\ket{a}+\ket{c}\otimes\ket{c}
\;=\; \bigl[\;\1\;\;\0\;\;\0\;\;\1\;\bigr]^\mathrm{T} \;,\cr
\ket{(bc)}
& = & \ket{a}\otimes\ket{a}+\ket{b}\otimes\ket{c}+\ket{c}\otimes\ket{b}
\;=\; \bigl[\;\1\;\;\1\;\;\1\;\;\0\;\bigr]^\mathrm{T} \;,\cr
\ket{(ca)}
& = & \ket{a}\otimes\ket{c}+\ket{b}\otimes\ket{b}+\ket{c}\otimes\ket{a}
\;=\; \bigl[\;\0\;\;\1\;\;\1\;\;\1\;\bigr]^\mathrm{T} \;,\cr
\ket{(abc)}
& = & \ket{a}\otimes\ket{b}+\ket{b}\otimes\ket{c}+\ket{c}\otimes\ket{a} 
\;=\; \bigl[\;\1\;\;\1\;\;\0\;\;\1\;\bigr]^\mathrm{T} \;,\cr
\ket{(acb)}
& = & \ket{a}\otimes\ket{c}+\ket{c}\otimes\ket{b}+\ket{b}\otimes\ket{a} 
\;=\; \bigl[\;\1\;\;\0\;\;\1\;\;\1\;\bigr]^\mathrm{T} \;,
\end{eqnarray}
where the state labels reflect the transformation properties of each state
under the group of global basis transformations in $V\otimes V$, 
which is isomorphic to the group of permutations of the three states labels $a$, $b$, and $c$.

\begin{figure}[b]
\includegraphics[width=12cm]{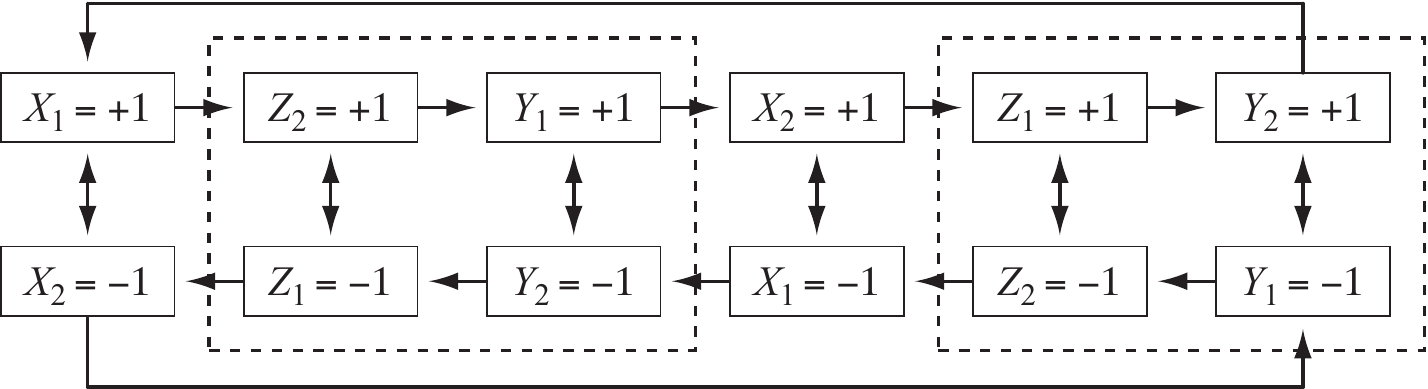}
\caption{Classical implications of the probabilities listed in Table~\ref{Probs}.}
\label{Implications}
\end{figure}

Products of the `spin' observables are defined as
\begin{equation}
A_{rs}A_{tu}
\,=\,\{
\,\bra{\bar{r}}\otimes\bra{\bar{t}}\,,
\,\bra{\bar{r}}\otimes\bra{\bar{u}}\,,
\,\bra{\bar{s}}\otimes\bra{\bar{t}}\,,
\,\bra{\bar{s}}\otimes\bra{\bar{u}}\,
\}\;,
\end{equation}
the four tensor products representing the outcomes
$++$, $+-$, $-+$, and $--$,
and the expectation value giving the correlation between the two `spins.' 
As an example, we calculate the probabilities and correlations 
of all product observables for the state $\ket{S}$ and obtain Table.~\ref{Probs}.
\begin{table}[t]
\begin{tabular}{|c||c|c|c|c||c|}
\hline
\ Observable\ \ &\ $++$\ \ &\ $+-$\ \ &\ $-+$\ \ &\ $--$\ \ &\ E.V. \ \\
\hline
$X_1X_2$, $Y_1Y_2$, $Z_1Z_2$ & $0$            & $\dfrac{1}{2}$ & $\dfrac{1}{2}$ & $0$            & $-1$ \phantom{\bigg|} \\
\hline
$X_1Y_2$, $Y_1Z_2$, $Z_1X_2$ & $\dfrac{1}{3}$ & $\dfrac{1}{3}$ & $0$            & $\dfrac{1}{3}$ & $+\dfrac{1}{3}$ \phantom{\bigg|}\\
\hline
$X_1Z_2$, $Z_1Y_2$, $Y_1X_2$ & $\dfrac{1}{3}$ & $0$            & $\dfrac{1}{3}$ & $\dfrac{1}{3}$ & $+\dfrac{1}{3}$ \phantom{\bigg|}\\
\hline
\end{tabular}
\caption{Probabilities and expectation values of product observables in the singlet state $\ket{S}$.
}
\label{Probs}
\end{table}
These probabilities cannot be reproduced 
in any classical hidden variable theory.
To see this, note that the first row of Table~\ref{Probs} indicates that
the `spin' pairs $(X_1X_2)$, $(Y_1Y_2)$, and $(Z_1Z_2)$ are all completely anti-correlated.
Row 2 indicates that if the first spin of the
pairs $(X_1Y_2)$, $(Y_1Z_2)$, or $(Z_1X_2)$ is $-1$, the second spin must also be $-1$,
whereas if the second spin is $+1$, the first spin must also be $+1$.
Row 3 indicates that if the first spin of the
pairs $(X_1Z_2)$, $(Z_1Y_2)$, or $(Y_1X_2)$ is $+1$, the second spin must also be $+1$,
whereas if the second spin is $-1$, the first spin must also be $-1$.
These implications are charted in the diagram shown in Fig.~\ref{Implications},
and it is not difficult to see that no classical configurations of spins can
satisfy all these conditions.  This argument is analogous to those of
Refs.~\cite{GHZ,GHSZ,Hardy:1993zza} for canonical QM.

To calculate the CHSH bound for our model, it suffices to examine all possible correlators
for the singlet state $\ket{S}$ only.
This is because all 6 entangled states can be transformed
into $\ket{S}$ via local basis transformations, that is, basis transformations in
only one of the tensored vector spaces.
We can also restrict the observables entering the correlator to just $X$, $Y$, and $Z$
and ignore their negatives since
\begin{equation}
\vev{A,a\,;B,b}
\,=\, \vev{A,-a\,;b,B}
\,=\, -\vev{-A,a\,;b,B} 
\,=\, \vev{a,A\,;B,-b}
\,=\, -\vev{a,A\,;-B,b}\,.
\end{equation}
Then, it is a fairly simple task to show that the CHSH bound for this model is 
the classical value of 2, despite the fact that its predictions cannot
be mimicked by any hidden variable theory.
It has been shown in Refs.~\cite{Chang:2012eh} and \cite{Chang:2012gg} that
this property holds for any Galois field $GF(q)$, not just for the case $q=2$,
as long as the prescriptions in this talk are followed.


We end this talk by pointing out that the above construction of a quantum-like theory on
vector spaces over $\K=GF(q)$ is not unique.
Instead of retaining the expression for probabilities, Eq.~(\ref{Pdef}),
from canonical QM, one can retain the following expression for expectation values instead:
\begin{equation}
\vev{A}_\psi\;=\;
\dfrac{\bra{\psi}\hat{A}\ket{\psi}}{\braket{\psi}{\psi}}\;.
\label{Avev}
\end{equation}
Due to the absence of an inner product in $V=\K^N$,
the application of this expression requires careful redefinitions 
of normalizable states, hermitian operators, and 
dual-vectors as hermitian conjugates of vectors.
One also needs to map $\bra{\psi}\hat{A}\ket{\psi}/\braket{\psi}{\psi}\in \K$
to a number in $\R$ if the operator is to represent
a physical observable.
These issues are addressed in detail in Ref.~\cite{Chang:2012xx}.
There, quantum-like theories are constructed on vector
spaces over $GF(3)$ and $GF(9)$, which are distinct from the
ones constructed here.
The CHSH bound for those theories are 2 for the $GF(3)$ case,
but 4 for the $GF(9)$ case.
This provides an existence proof of quantum-like theories
that violate the Cirel'son bound, while at the same time
indicating that the replacement of $\C$ with $GF(q)$
does not lead to a unique CHSH bound.
Rather, it is the absence of the inner product and
how it is dealt with that is apparently responsible for the difference.
For the case in which Eq.~(\ref{Avev}) is retained,
the CHSH bound also depends on whether $GF(q)$ has a complex-like structure or not \cite{Chang:2012xx}.

One thing that is clear from our work is that the existence of an inner product is not
a necessary `gene' for the survival of a quantum theory.
This discovery opens up the possibility of constructing
quantum-like theories on other vector spaces that have heretofore not
been considered.
Further investigations into these matters will be reported 
in future publications.


\bigskip
{\bf Acknowledgments:}
We would like to thank Sir Anthony Leggett and Prof. Chia Tze for helpful discussions.
ZL, DM, and TT are supported in part by
the U.S. Department of Energy, grant DE-FG05-92ER40677, task A.

\bibliographystyle{aipproc}   
\bibliography{QTRF6-takeuchi}

\end{document}